\documentclass[a4]{article}

\usepackage{tikz}
\usetikzlibrary{shapes}
\usetikzlibrary{positioning}
\usetikzlibrary{arrows}
\usetikzlibrary{calc}
\usetikzlibrary{shadows}

\usepackage{listings}
\lstdefinelanguage{x10}
{
	keywords = {
		abstract, as, assert,
		async, at, ateach,
		atomic, break, case,
		catch, class, clocked,
		continue, def, default,
		do, else, extends,
		false, final, finally,
		finish, for, goto,
		haszero, if,
		implements, import, in,
		instanceof, interface, native,
		new, null,
		offers, operator,
		package, private, property,
		protected, public,
		return, self, static,
		struct, super, switch,
		this, throw, transient,
		true, try, type,
		val, var, void,
		when, while
	},
	morekeywords = {
		atomic,
		here,
		next,
		offer,
		resume
	},
	comment = [l]{//},
	morecomment = [s]{/*}{*/},
	string = [b]",
}

\usepackage{xcolor}
\usepackage{fixltx2e}

\usepackage{hyperref}

\def\ilet{\emph{i}-let}
\def\ilets{\emph{i}-lets}
\def\ipos{\emph{i}POS}
\def\irts{\emph{i}RTSS}

\def\cic{C\emph{i}C}
\def\ipos{{OctoPOS}}

\hypersetup{
   bookmarks,
   colorlinks=true,
	linkcolor=black,
	citecolor=black,
	urlcolor=black,
}

\tikzset{
  /tikz/myshadow/.style={
    general shadow={%
      shadow scale=1,
      shadow xshift=.03ex,
      shadow yshift=-.03ex,
      opacity=1,
      fill=black!4,
      every shadow,
      #1,
      shadow xshift/.expanded=22*\pgfkeysvalueof{/tikz/shadow xshift},
      shadow yshift/.expanded=22*\pgfkeysvalueof{/tikz/shadow yshift}
    },
    general shadow={%
      shadow scale=1,
      shadow xshift=.03ex,
      shadow yshift=-.03ex,
      opacity=1,
      fill=black!6,
      every shadow,
      #1,
      shadow xshift/.expanded=20*\pgfkeysvalueof{/tikz/shadow xshift},
      shadow yshift/.expanded=20*\pgfkeysvalueof{/tikz/shadow yshift}
    },
    general shadow={%
      shadow scale=1,
      shadow xshift=.03ex,
      shadow yshift=-.03ex,
      opacity=1,
      fill=black!8,
      every shadow,
      #1,
      shadow xshift/.expanded=18*\pgfkeysvalueof{/tikz/shadow xshift},
      shadow yshift/.expanded=18*\pgfkeysvalueof{/tikz/shadow yshift}
    },
    general shadow={%
      shadow scale=1,
      shadow xshift=.03ex,
      shadow yshift=-.03ex,
      opacity=1,
      fill=black!12,
      every shadow,
      #1,
      shadow xshift/.expanded=16*\pgfkeysvalueof{/tikz/shadow xshift},
      shadow yshift/.expanded=16*\pgfkeysvalueof{/tikz/shadow yshift}
    },
    general shadow={%
      shadow scale=1,
      shadow xshift=.03ex,
      shadow yshift=-.03ex,
      opacity=1,
      fill=black!20,
      every shadow,
      #1,
      shadow xshift/.expanded=14*\pgfkeysvalueof{/tikz/shadow xshift},
      shadow yshift/.expanded=14*\pgfkeysvalueof{/tikz/shadow yshift}
    },
    general shadow={%
      shadow scale=1,
      shadow xshift=.03ex,
      shadow yshift=-.03ex,
      opacity=1,
      fill=black!26,
      every shadow,
      #1,
      shadow xshift/.expanded=12*\pgfkeysvalueof{/tikz/shadow xshift},
      shadow yshift/.expanded=12*\pgfkeysvalueof{/tikz/shadow yshift}
    },
    general shadow={%
      shadow scale=1,
      shadow xshift=.03ex,
      shadow yshift=-.03ex,
      opacity=1,
      fill=black!30,
      every shadow,
      #1,
      shadow xshift/.expanded=10*\pgfkeysvalueof{/tikz/shadow xshift},
      shadow yshift/.expanded=10*\pgfkeysvalueof{/tikz/shadow yshift}
    },
    general shadow={%
      shadow scale=1,
      shadow xshift=.03ex,
      shadow yshift=-.03ex,
      opacity=1,
      fill=black!34,
      every shadow,
      #1,
      shadow xshift/.expanded=8*\pgfkeysvalueof{/tikz/shadow xshift},
      shadow yshift/.expanded=8*\pgfkeysvalueof{/tikz/shadow yshift}
    },
    general shadow={%
      shadow scale=1,
      shadow xshift=.03ex,
      shadow yshift=-.03ex,
      opacity=1,
      fill=black!38,
      every shadow,
      #1,
      shadow xshift/.expanded=6*\pgfkeysvalueof{/tikz/shadow xshift},
      shadow yshift/.expanded=6*\pgfkeysvalueof{/tikz/shadow yshift}
    },
    general shadow={%
      shadow scale=1,
      shadow xshift=.14ex,
      shadow yshift=-.14ex,
      opacity=1,
      fill=black!40,
      every shadow,
      #1
    }
  }
}

\begin{document}

\title{{\bf Invasive Computing - Common Terms and Granularity of Invasion}\thanks{Dagstuhl Seminar
    13052, ``Multicore Enablement for Embedded and Cyber Physical
    Systems'', organizers: Andreas Herkersdorf, Michael G. Hinchey,
    Michael Paulitsch, 27.01.13--01.02.13}}

\author{J{\"u}rgen Teich, Wolfgang Schr{\"o}der-Preikschat, Andreas Herkersdorf\\
          DFG Transregional Collaborative Research Centre 89\\
          Invasive Computing\\
          {\tt www.invasic.de}
}

\maketitle

\begin{abstract}
Future MPSoCs with 1000 or more
processor cores on a chip require new means for  \emph{resource-aware programming} in order to
deal with increasing imperfections such as process variation, fault rates, aging
effects, and power as well as thermal problems. On the other hand, predictable 
program executions are threatened if not impossible if no proper means of resource 
isolation and exclusive use may be established on demand. In view of these problems and menaces, 
\emph{invasive computing} enables an application programmer to claim 
for processing resources and spread computations to claimed processors
dynamically at certain points of the program execution.
Such decisions may be depending on the degree of application parallelism and the state of the underlying resources such as utilization, load, and temperature, but also with the goal to 
provide predictable program execution on MPSoCs by claiming processing resources exclusively as the 
default and thus eliminating interferences and creating the necessary isolation between
multiple concurrently running applications. For achieving this goal, invasive computing 
introduces new programming constructs for resource-aware programming that meanwhile, 
for testing purpose, have been embedded into the parallel computing language X10 as developed by IBM
using a library-based approach.
This paper presents major ideas and common terms of invasive computing as investigated 
by the DFG Transregional Collaborative Research Centre TR89. Moreoever, a reflection 
is given on the granularity of resources that may be requested by invasive programs. 
\end{abstract}

\section*{Invasive Computing - An Overview}
With the ever increasing number of cores that may be integrated on a single chip, difficulties arise when programming SoC devices in a resource-efficient manner. Also, the predictability of non-functional 
properties of program execution such as performance, safety and security is 
hopeless if no means for separation and elimination of information flow interferences caused 
by multiple programs sharing the resources on the MPSoC may 
be established and guaranteed during program execution.
We see invasive computing \cite{Tei11} as a solution to the above problems by
envisioning that applications running on Multi-Pro\-ces\-sor System-on-a-Chip architectures (MPSoC) may request and distribute their workload themselves based on their temporal computing demands, temporal availability of resources, and other state information of the resources (e.\,g., temperature, faultiness, resource usage, permissions).

However, in order to make this computing paradigm become a reality and to evaluate its benefits properly, the way of application development including algorithm design, language implementation and compilation tools needs to change to a large extent.

On the one hand, the idea of allowing applications to spread their computations on 
claimed resources and later free them again sounds promising.
The expected benefits include an increase of speedup (with respect to
statically mapped applications), fault-tolerance, and a considerable
increase of resource utilization, hence computational efficiency.  These
efficiency numbers, however, need to be analyzed carefully and traded against the overhead caused
with respect to statically mapped applications.
However, and more importantly, being able to claim the exclusive access to sets of processing, memory 
and communication resources during execution time frames shall allow to make 
multi-core program execution much more predictable with respect to non-functional properties 
such as execution time, safety and security properties.

First and most fundamentally, in \cite{Tei08b} and \cite{Tei11}, Teich and others introduced the paradigm of  \emph{invasive computing} that integrates research on algorithm and program 
design as well as micro- and macro-architectural changes of MPSoCs to support invasive 
programming.
The main idea of \emph{invasion} is to add to a parallel program the ability to explore and claim resources in a certain neighborhood and to copy its program and possibly data to such places 
in a phase of invasion, and then to execute the given problem in parallel based on the available (invasible) region of processing resources.
Through invasion, an application will thus be able to spread its computations for parallel execution based on the availability and the actual state of processing resources.
For execution phases of reduced degree of available application parallelism, the application may itself perform a \emph{retreat} to free occupied resources so to optimally exploit all resources and make them available for other applications.

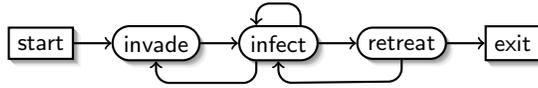
\begin{figure}[bt]
\vspace{-1em}
\centering
\begin{tikzpicture}[
	endp/.style={
	    draw,
	    rectangle,
	    font=\sffamily \small,
	    thick,
            /tikz/myshadow,
            fill=white
	},
	state/.style={
	    draw,
	    rounded rectangle,
	    font=\sffamily \small,
	    thick,
            /tikz/myshadow,
            fill=white
	},
	arrow/.style={
	    ->,
	    rounded corners=5pt,
	    thick
	},
	scale=.51]

    \node [endp] (start) {start};
    \node [state, right=5mm of start] (invade) {invade};
    \node [state, right=5mm of invade] (infect) {infect};
    \node [state, right=5mm of infect] (retreat) {retreat};
    \node [endp, right=5mm of retreat] (exit) {exit};
    \draw [arrow] (start.east) -- (invade.west);
    \draw [arrow] (invade.east) -- (infect.west);
    \draw [arrow] (infect.east) -- (retreat.west);
    \draw [arrow] (retreat.east) -- (exit.west);
    \draw [arrow] (infect.north east) -- ++(0,16pt) -- ($(infect.north west) + (0, 16pt)$) -- (infect.north west);
    \draw [arrow] (infect.south west) -- ++(0,-16pt) -- ($(invade.south) + (0,-16pt)$) -- (invade.south);
    \draw [arrow] (retreat.south) -- ++(0,-16pt) -- ($(infect.south) + (0,-16pt)$) -- (infect.south);
\end{tikzpicture}
\vspace{-0.5\baselineskip}
\caption{State chart of an invasive program.}
\label{fig:statechart}
\vspace{-\baselineskip}
\end{figure}

The chart depicted in \autoref{fig:statechart} shows the typical state transitions that occur during the execution of an invasive program.
In the beginning, an initial \emph{claim} has to be constructed.
By \emph{claim} we denote a set of processor resources that the application 
can use for its parallel execution.
Claim construction is done by issuing a call to \emph{invade}.
After that, \emph{infect} is used to start the actual application code on the previously allocated \emph{claim}.
The actual application code that is spread onto infected resources for subsequent parallel execution is called \emph{i-let} (and will be explained in the following Section Common Terms).
Once the execution on all claimed cores finishes, the number of cores inside the \emph{claim} can be altered by calling \emph{invade} or \emph{retreat} to either expand or shrink the application's \emph{claim}.
In case of \emph{retreat}, the processing elements are cleaned up from the \emph{i-let} entities that have been setup by \emph{infect}.
Alternatively, if the degree of parallelism does not change, it is also feasible to dispatch a different program onto the same set of cores by issuing another call to \emph{infect}.
If a call to \emph{retreat} leaves the \emph{claim} empty, there are no computing resources left for further execution of the program, hence it terminates its execution and exits.
Notably, a claim may not only contain processing resources, but also memory as well as communication resources.\\[1.5mm]
We do believe that invasive computing might solve many future problems of massively parallel application processing on future MPSoC platforms by providing and porting principles of \emph{self-organization} into reconfigurable architectures, integrating 1000 and more processor cores on a single chip.
Moreover, better predictable timing, enablement of safety-properties on demand, and respecting 
security properties will be result by the exclusive nature of a claim providing isolation 
of program and data as a must and reducing interferences between multiple 
concurrently running programs to a minimum. 
Other major advantages of invasive computing are given by the feature of resource-awareness, a gain in computational efficiency and performance, application-level error resiliency, self-adaptive power control and management, and self-optimization of resource utilization.
Another final objective is to increase the lifetime or to encompass aging effects of future sub-micron technology by avoiding stressing the hardware too much.

In the following, important common terms of invasive computing as investigated currently 
within the equally named DFG collaborative research centre TR89 are discussed.

\section*{Common Terms}
A piece of program subjected to parallel processing according to the paradigm of invasive computing is referred to as an ``invasive-let'': in short \ilet\footnote{This conception goes back to the notion of a ``servlet'', which is a (Java) application program snippet target for execution within a web server.}. Depending on the level of abstraction considered, different \ilet{} entities and associated properties are distinguished:
\begin{description}
\item[candidate] (a) prospect out of a family of algorithms for the same problem to be solved, (b) potential cause of a specific operating mode of the (parallel) processor as to be enforced by \irts\footnote{\irts{} is the acronym of our 
run-time system for invasive MPSoCs.} and (c) possibly represented and maintained as a separate source module.
\item[instance] (a) medium of activity of an invasive-parallel program, (b) specification of a virtual processor for it and (c) possibly represented and maintained as a separate object module.
\item[incarnation] (a) characteristic of the mode of operation to be realised by \irts, (b) ground anchor for the resources virtually needed for making progress in parallel processing and (c) possibly represented and maintained as a separate load module.
\item[execution] (a) actual disposition of a portion of an invasive-parallel program running on a real processor, (b) effective unit of processing implemented in soft-, firm-, or hardware (c) associated with a dedicated memory image.
\end{description}
Given these notions of \ilet{s} and taking an operating system's  point of view, candidates and instances are user-level entities while incarnations and executions are system-level entities. At system level, two more terms have been established which manifest in corresponding \irts{}  abstractions:
\begin{description}
\item[claim] designates a particular set of hardware resources made available to an invading process on demand and according to selected constraints.
\item[team] designates a particular set of \ilet{} entities (i.e., incarnations) associated with a specific claim.
\end{description}
These two abstractions aid application-level processes in the description of (static/dynamic) resource demands, the indication of the operating mode of the computing machine and the modelling of a certain run-time behaviour of the constituting \ilets.

Discussions within the collaborative research centre revealed that the generally usual notion of ``application'' has quite different meanings in the diverse technical disciplines. The range goes from a single ``thread'' within a (non-sequential, multi-threaded) program looking into a very dedicated task to a (possibly complex formation of a) logically self-contained assembly of programs that jointly performs a certain computation or control function. By way of example, the former case relates to read-out of a sensor device and the latter case to some feedback control system consisting of many sensors, actuators, and (hardware/software) means for human-computer interaction. Within the collaborative research centre TR89, ``application'' much more corresponds to the latter than the former.

\section*{Granularity of Invasion and Infection}
MPSoCs containing 100s of processors will be typically organized in groups called \emph{tiles}.
Whereas inside a tile of processors, shared memory communication is possible, the communication
between tiles is organized by message passing and supported typically by one or more 
\emph{networks-on-a-chip} (NoC(s)).
The question of the adequate granularity of invasion---namely core or tile---of invasion, as significant at \texttt{invade}-time, \emph{and} infection, as significant at \texttt{infect}-time, in terms of the hardware units affected by the respective measures is a central issue of the TR89. These two actions while executing an invasive (parallel) program establish the moment of \emph{allocation} of hardware units requested by an application (entity) and \emph{dispatching} of \ilets{} to some processing element (i.\,e., core), respectively. A common understanding was to differentiate between these moments (``separation of concerns'') and, as a further consequence, to accept different granularity depending on the level of abstraction considered.

\emph{Granularity of invasion} is interlinked with the \emph{guarantees} the hard- and software system has to give to applications. This depends on (1) the resource-allocation constraints of the application specified by \texttt{invade}, (2) the scheduling criteria implemented by the \irts{} and (3) the assertiveness of the system-software/hardware stack to enforce the claimed constraints. Several artefacts of the system software and the hardware may be the cause of failure to comply with the set of \emph{constraints} that may stated by an application
as parameters to an \texttt{invade}
call. As an example, such a constraint may be: ``Provide me an exclusive claim of four cores
all belonging to the same tile". Besides temporal unavailability of a certain hardware unit (e.\,g., due to overheating or transient errors), typical cases of such artefacts come with coordinated sharing of system-level (hard-/software) resources such as cores, caches, busses or memory, and with the \emph{interference} of otherwise unrelated application processes. Further anomalies may arise through the kind of (process) scheduling criteria that form the basis of design and implementation of (parts of) an operating system. Here, \emph{user-oriented criteria} (e.\,g., response time, cycle time) are in opposition to \emph{system-oriented criteria} (e.\,g., utilisation). The latter imply potential hazard to applications that assume a predictable run-time behaviour of the underlying computing system; they are typical of general-purpose systems. The former are likely to let hardware resources rest in favour of deterministic operation, they are typical for a special-purpose system. Being in charge of juggling with both kinds of scheduling criteria at the same time (in practice within an operating system) means, however, to give priority to either of them. This breeds interference of the other, respectively. \emph{As predictable run-time behaviour is an important aspect of \emph{InvasIC}\footnote{Acronym of the DFG TR89, see \cite{TR89web} 
for more details.}, in \irts{}, user-oriented criteria dominate system-oriented criteria.
This is reflected by an API that demands the specification of (mandatory/optional) constraints from an application process in order to claim (i.\,e., \texttt{invade}) hardware resources.}

For \irts{}, the meaning of constraints is two-fold and distinguishes mandatory from optional specifications on the part of a particular application process. \emph{Mandatory constraints} of invasion declare the resource demands of an imminent computation phase and provide an indication of the expected benefit of resource allocation, in functional and non-functional terms.
\emph{Optional constraints} qualify the willingness to share the allocated resources with competing processes of other (unrelated) applications, in spatial and temporal terms, and notify toleration of temporary under-/oversupply of spare cores for \ilet{} dispatching. The former are for the \emph{quantification} of application requirements, while the latter are for \emph{immunisation} of (parts of) an application.
\emph{By default, resources are exclusively allocated to applications, but the exclusiveness may gradually be loosened by way of optional constraints.}

Throughout the last year, the focus of resource allocation was on physical processing elements such as cores or tiles (of cores), respectively. But note that this focus also depends on the position taken within a multi-layer computing system such as considered in the DFG TR89. At a lower (i.\,e., more hardware-oriented) level of abstraction, \ipos{}\footnote{Acronym of the operating system of an invasive MPSoC.} operates in a coarse-grained manner and allocates tiles to an agent system upon request. On a higher (i.\,e., more application-oriented) level, the agent system works in a fine-grained fashion and allocates the cores of one or more tiles to the (C/C++, X10) run-time system upon request. Such an approach of task sharing in resource management is very common in today's computing systems and has proved itself. Thus, core granularity of resource allocation is seen at application level even though tile granularity forms the basis on a lower level within the system.

An important influencing factor on the granularity of resource allocation is given with the (optional) constraint of application immunisation as mentioned above. Assume that an application wants to exclusively use a compute tile in order to avert interference by some other application as far as possible. In such a situation, which reflects the default case, core allocation to the latter application always starts from a ``virgin'' tile even if the (last) tile that was allocated to the former has one or more cores to spare. That is to say, \irts{}  tolerates \emph{internal fragmentation} of a tile for the benefit of a more predictable run-time behaviour. As a consequence, this means a tile granularity of resource allocation, namely to assure immunisation of (parts of) an application.
\emph{Thus, tile granularity will be the (default, but overridable) praxis although core granularity is logically seen at application level.}

\emph{Granularity of infection} largely depends on the nature and configuration of the claim of hardware that is going to be \texttt{infect}-ed by (a team of) \ilets{} in order to initiate a parallel computation. At that point in time, \irts{} (more specifically, \ipos{}) deploys \ilet{} incarnations with the aid of the \cic{}\footnote{\cic{} is the acronym for core \ilet{} controller, 
a hardware unit serving as an \ilet{} dispatcher on a tile of processors.}. At the lowest (i.\,e., hardware) level of abstraction, the \ilet{} dispatching according to the constraints of the team's claim always takes place at a core granularity. The \cic{} makes its (rule-based) dispatching decisions on the basis of the claim identification associated with the deployed \ilets{}. Only in case of an \ilet{} tagged with a ``wildcard'' identifier (\texttt{null}) will the \cic{} select any core of the compute tile, adhering to system-oriented optimisation criteria (such as utilisation) for tile-wide load balancing at \ilet{} arrival time. In case of a valid (``non-\texttt{null}'') claim identifier, however, the \cic{} first and foremost adheres to user-oriented optimisation criteria (such as response or cycle time), and dispatches the \ilets{} to the cores of the tile as constrained by that very identifier. That is to say, if resource allocation---by means of \texttt{invade} and overriding the system default of exclusive use---resulted in the sharing of a single compute tile amongst (entities of) different applications, the \cic{} will send \ilets{} only to those cores that belong to the claim of the respective \ilet{}. In that case, system-oriented optimisation criteria come after user-oriented ones, if at all.

This claim-based differentiation is made for better control of \emph{interference} in case of multi-programmed compute tiles that are claimed (i.\,e., shared) by applications of different and possibly conflicting quality requirements in terms of non-functional properties (such as timing, jitter, energy or noise). In the process of setting out a claim (\texttt{invade}), the agent system of \irts{} establishes the appendant \cic{} dispatching rule that later on gets activated by \ipos{} in the process of \ilet{} deployment (\texttt{infect}). When a team of \ilets{} is assorted for a specific claim---after return from a successful call to \texttt{invade}, but before the call to \texttt{infect} for that very claim---the association between \ilet{} and claim identifier or wildcard, respectively, is established. During infection, \ipos{} then tags all \ilets{} with the identifying information related to the claim of their team.

For the purpose of better system utilisation, the \cic{} will be capable of dispatching \ilets{} of an application to spare cores even of an exclusively taken compute tile that, however, was not entirely allocated to the application. The number of spare cores then corresponds to the portion of internal fragmentation (of such a tile) as result of application immunisation as explained above. Utilisation of these cores then leads to a temporary oversupply of computing resources to the application running on the respective compute tile. This also brings about interference and causes unpredictable run-time behaviour of an application. Just like oversupply, also a temporary undersupply of computing resources may occur. An example of this is an over-heated core that will be masked by the \cic{} and, thus, excluded from further \ilet{} processing until its operating temperature has dropped below a certain threshold. Both over- and undersupply affect application processing in non-functional terms. By default, \irts{} will not instruct the \cic{} to oversupply an application with spare cores, but this presetting may be overridden by means of optional constraints specified by an application (at \texttt{invade}-time). The same goes for the undersupply of (computing) resources, which is also considered an optional constraint of invasion to give application-side toleration notice to \irts{}.

\section*{Acknowledgments}
This work was supported by the German Research Foundation (DFG) as part of the Transregional Collaborative Research Centre  ``Invasive Computing" \\
(SFB/TR 89).

%
\bibliographystyle{abbrv}
\bibliography{DAGstuhl13052}

\begin{thebibliography}{1}

\bibitem{TR89web}
{DFG Transregional Research Centre 89}.
\newblock {Invasive Computing}.
\newblock {\tt http://www.invasic.de}.

\bibitem{Tei08b}
J.~{Teich}.
\newblock Invasive {Algorithms} and {Architectures}.
\newblock {\em it - Information Technology}, 50(5):300--310, 2008.

\bibitem{Tei11}
J.~Teich, J.~Henkel, A.~Herkersdorf, D.~Schmitt-Landsiedel,
  W.~Schr{\"o}der-Preikschat, and G.~Snelting.
\newblock {\em Multiprocessor {System-on-Chip}: {Hardware} {Design} and {Tool}
  {Integration}}, chapter 11, Invasive {Computing}: {An} {Overview}, pages
  241--268.
\newblock Springer, 2011.

\end{thebibliography}

\end{document}